\documentclass[12pt,aps,prd,superscriptaddress,floatfix]{article}
\usepackage{cite}
\usepackage[bookmarks=false]{hyperref}
\hypersetup{pdfstartview={XYZ null null 1.00}}
\usepackage[left=2.5cm,right=1cm,top=2cm,bottom=2cm,bindingoffset=0cm]{geometry}
\usepackage{euscript}
\usepackage{amsfonts}
\usepackage{amsmath}
\usepackage{amssymb}
\usepackage{dsfont}
\usepackage{longtable}
\usepackage{tabularx}
\usepackage{multirow}
\usepackage[dvips,pdftex]{graphicx}
\graphicspath{{pics/}}

\newcommand{\be}{\begin{equation}}
\newcommand{\ee}{\end{equation}}
\newcommand{\bes}{\begin{equation*}}
\newcommand{\ees}{\end{equation*}}
\newcommand{\bdis}{\begin{displaymath}}
\newcommand{\edis}{\end{displaymath}}
\newcommand{\bga}{\begin{equation}\begin{gathered}}
\newcommand{\ega}{\end{gathered}\end{equation}}
\newcommand{\bgas}{\begin{equation*}\begin{gathered}}
\newcommand{\egas}{\end{gathered}\end{equation*}}
   
\newcommand{\Tr}{\mathop{\mathrm{Tr}}\nolimits}

\DeclareMathOperator\arctanh{arctanh}
\DeclareMathOperator\diag{diag}

\begin{document}
\title{Resummation of relativistic corrections to heavy $\text{quarkonium}+\gamma$ production at $Z$ factory}
\author{Anton Trunin\thanks{amtrnn@gmail.com}}
\date{}
\maketitle
\begin{abstract}
Cross sections of the heavy quarkonium production process $e^+e^-\to H + \gamma$ at the c.m. energy of $Z^0$-boson resonance are calculated at leading order in $\alpha_s$ with relativistic corrections resummed within the framework of nonrelativistic QCD (NRQCD). $S$- and $P$-wave charmonia and bottomonia are considered. For $J/\psi(\Upsilon)$ production, the result of vector-meson-dominance (VMD) approach is also given. Relativistic contributions to charmonium production are of the same magnitude as next-to-leading order QCD corrections, decreasing cross section 20--30\%. Relativistic corrections to bottomonium production are negligible, not exceeding 2\%.

\end{abstract}

\section{Introduction}
The prospects of high-luminosity electron--positron collider operating at the energy of $Z^0$-boson mass, often referred to as (super) $Z$ factory or GigaZ, are widely discussed~\cite{erler2000,aguilar-saavedra2001,aarons2007,ma2010,dong2018,fujii2019,agapov2022}. With projected luminosity $\mathcal L=10^{34}$--$10^{36}\,\text{cm}^{-2}\,\text{s}^{-1}$, this experiment allows study of single heavy quarkonium~$H$ production, with the significance of $Z^0$-boson production channel, as opposite to the similar processes observed at $B$ factories. Till the recent time, the theoretical description of $e^+e^-\to H+\gamma$ processes at $\sqrt s = M_Z$ were performed at leading (LO) and next-to-leading (NLO) orders in strong coupling constant $\alpha_s$ and at zero (i.e., nonrelativistic) order in heavy quark relative velocity $v^2$~\cite{chang2010,chen2013,chen2014,sun2014}. Still, as it is well-known from NRQCD power-counting rules~\cite{nrqcd-rules, nrqcd}, relativistic corrections in $v^2$ are of the same order as those in $\alpha_s$ and have also to be dealt with.

In this paper we calculate relativistic corrections to the color-singlet heavy quarkonium production process $e^+e^-\to H+\gamma$ at LO in $\alpha_s$ employing resummation approach of NRQCD~\cite{bodwin-resum}.
We cover $S$- and $P$-wave states and compare results with NLO $\alpha_s$ corrections calculated in Refs.~\cite{chen2014,sun2014} for charmonium (though, only partially for $J/\psi$) and bottomonium, respectively.
For $J/\psi$ and $\Upsilon$ case we also use vector-meson-dominance (VMD) approach~\cite{bodwin-vmd} to effectively include all $\alpha_s$ and $v^2$ corrections in the photon fragmentation channel.
The processes in consideration were also studied in two more recent papers~\cite{chung19,liao22}. Ref.~\cite{chung19} considers $\eta_{c,b}+\gamma$ production at NLO in $\alpha_s$ and at next-to-leading-logarithm in ratio of quark mass to c.m. energy, including $v^2$ relativistic corrections as well. Ref.~\cite{liao22} gives nonrelativistic analysis at LO $\alpha_s$ of excited charmonium and bottomonium production up to $4S$- and $4P$-wave states.

This paper is organized as follows. In Section~\ref{sec:nrqcd_resum} brief overview of NRQCD resummation approach is given. Section~\ref{sec:amplitude} introduces the perturbative amplitude for heavy quark--antiquark $\text{pair}+\gamma$ production at LO in $\alpha_s$, while the issue of its angular integration is addressed in Section~\ref{sec:ang-int}. Section~\ref{sec:jpsi} is devoted to the cross section of $e^+e^-\to J/\psi(\Upsilon)+\gamma$ process, in both resummed NRQCD and VMD approaches. Section~\ref{sec:numerical} contains numerical results for $S$- and $P$-wave charmonia and bottomonia, comparison of relativistic results with nonrelativistic approximation and with NLO $\alpha_s$ contributions, and cross section distributions. Analytical expressions for differential and integrated cross sections are explicitly given in Appendix.

\section{NRQCD amplitude and resummation approach}\label{sec:nrqcd_resum}
The NRQCD amplitude for the production process $e^+e^-\to H+\gamma$ of heavy quarkonium~$H$ can be approximated to the following form~\cite{bodwin-resum}:
\bga
\label{eq:nrqcd-gen}
\mathcal M_\text{NRQCD}[e^+e^-\to H+\gamma]=\sqrt{2M} \,\langle H|\mathcal O|0\rangle\sum_{n=0}^\infty c_n\, {\langle {\boldsymbol p}^{2n} \rangle}_H,\\
\mathcal O=\mathcal P_0,\quad {\langle {\boldsymbol p}^{2n} \rangle}_H=\dfrac{\langle H|\mathcal P_n|0\rangle}{\langle H|\mathcal O|0\rangle},
\ega
where $M$ is hadron mass. 
The NRQCD operators $\mathcal P_n$ create color-singlet quark-antiquark pair $q\bar q$ with the same quantum numbers $^{2S+1}L_J$ as of quarkonium state~$H$:
\bga
\label{eq:nrqcd-operators}
\mathcal P_n(^1S_0)=\psi^\dagger\Bigl(\frac{-i}{2}\overset\leftrightarrow{\mathbf D}\Bigr)^{2n}\chi,\\
\mathcal P_n(^3S_1)=\psi^\dagger\Bigl(\frac{-i}{2}\overset\leftrightarrow{\mathbf D}\Bigr)^{2n}\boldsymbol\sigma\,\chi,\\
\mathcal P_n(^1P_1)=\psi^\dagger\Bigl(\frac{-i}{2}\overset\leftrightarrow{\mathbf D}\Bigr)^{2n+1}\chi,\\
\mathcal P_n(^3P_0)=\frac13\psi^\dagger\Bigl(\frac{-i}{2}\overset\leftrightarrow{\mathbf D}\Bigr)^{2n}
\Bigl(\frac{-i}{2}\overset\leftrightarrow{\mathbf D}\cdot\boldsymbol\sigma\Bigr)\chi,\\
\mathcal P_n(^3P_1)=\frac12\psi^\dagger\Bigl(\frac{-i}{2}\overset\leftrightarrow{\mathbf D}\Bigr)^{2n}
\Bigl(\frac{-i}{2}\overset\leftrightarrow{\mathbf D}\times\boldsymbol\sigma\Bigr)\chi,\\
\mathcal P_n(^3P_2)=\psi^\dagger\Bigl(\frac{-i}{2}\overset\leftrightarrow{\mathbf D}\Bigr)^{2n}
\Bigl(\frac{-i}{2}\overset\leftrightarrow{D}{}^{(i}\sigma^{j)}\Bigr)\chi,\\
\ega
where $\psi^\dagger(\chi)$ is (anti)quark creation spinor, $\boldsymbol\sigma$ are Pauli matrices, and $\overset\leftrightarrow{\mathbf D}=\overset\rightarrow{\mathbf D}-\overset\leftarrow{\mathbf D}$ is gauge-covariant derivative~\cite{nrqcd}.
The short-distance coefficients~$c_n$ in Eq.~\eqref{eq:nrqcd-gen} can be obtained by matching the amplitude of the process $e^+e^-\to q\bar q+\gamma$ calculated in conventional perturbative QCD at each order in squared relative momenta $\boldsymbol p^2$ of quark-antiquark pair $q\bar q$.
\be
\label{eq:nrqcd-short}
c_n=
\frac1{\sqrt{2N_c}}\frac1{n!}\frac{\partial^n}{\partial\boldsymbol p^{2n}}
\left(
\frac{|\boldsymbol p|^{-L}}{2\epsilon_{\boldsymbol p}}{\mathcal M_\text{pQCD}[e^+e^-\to q\bar q+\gamma]}\right)\biggr|_{\boldsymbol p^{2}=0}.
\ee
The factors $\sqrt{2N_c}$ and $2\epsilon_{\boldsymbol p}$ containing number of colors~$N_c$ and heavy quark energy~$\epsilon_{\boldsymbol p}=\sqrt{m^2+\boldsymbol p^2}$ account for normalization of NRQCD matrix elements and pQCD amplitude in~\eqref{eq:nrqcd-short}. According to the power-counting rules~\cite{nrqcd-rules}, operators $\mathcal P_{n>0}$ scale in heavy quark velocity as $v^{2n}$ relatively to $\mathcal O$, so that the first term $c_0{\langle H|\mathcal O|0\rangle}$ in Eq.~\eqref{eq:nrqcd-gen} can be considered as non-relativistic approximation to the amplitude, while the other terms $c_n{\langle H|\mathcal P_n|0\rangle}$ represent relativistic corrections to it. In Ref.~\cite{bodwin-resum} the resummation method for such corrections were presented on the basis of the relation~\cite{bodwin-gk}
\be
\label{eq:nrqcd-ggkr}
{\langle {\boldsymbol p}^{2n} \rangle}_H={\langle {\boldsymbol p}^{2} \rangle}_H^n
\ee
allowing to express matrix elements of the higher-order NRQCD operators~\eqref{eq:nrqcd-operators} through the single relativistic parameter~${\langle {\boldsymbol p}^{2} \rangle}_H$.
Then, the amplitude~\eqref{eq:nrqcd-gen} finally reads as
\be
\label{eq:nrqcd-amp}
\mathcal M_\text{NRQCD}[e^+e^-\to H+\gamma]=\frac{\sqrt{M}\,\langle H|\mathcal O|0\rangle}{2\sqrt{N_c}\,\epsilon_{\boldsymbol p}}\, {|\boldsymbol p|^{-L}}\mathcal M_\text{pQCD} \biggr|_{\boldsymbol p^{2}={\langle {\boldsymbol p}^{2} \rangle}_H}.
\ee
The generalized Gremm--Kapustin relation~\eqref{eq:nrqcd-ggkr} is valid up to the corrections of relative order $O(v^2)$~\cite{bodwin-resum}.
It means that resummed expression~\eqref{eq:nrqcd-amp} is accurate only up to the relative $O(v^4)$ corrections and so is formally equivalent, in the NRQCD formalism, to the original amplitude~\eqref{eq:nrqcd-gen} truncated at the $n=1$ term.
Still, the resummation method based on the relation~\eqref{eq:nrqcd-ggkr} is widely applied, since it allows to estimate the convergence rate of NRQCD expansion in~\eqref{eq:nrqcd-gen} and also, in principle, to present results of the calculation in closed analytical form~\cite{bodwin09,lee,sang,fan,bodwin2014}.
In the following we directly compare numerical predictions from the resummed cross sections and from the corresponding expressions with only leading correction in ${\langle {\boldsymbol p}^{2} \rangle}_H$ retained. The latter can be readily extracted from the closed form result for $\sigma[e^+e^-\to H+\gamma]$ supplied by Eq.~\eqref{eq:nrqcd-amp}.

\begin{figure}[t!]
\center\includegraphics{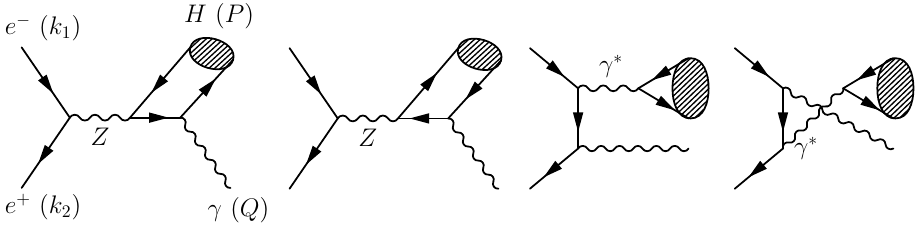}
\caption{The four leading order diagrams contributing to $e^+e^-\to H+\gamma$ via virtual $Z^0$-boson and photon $\gamma^*$.
The last two diagrams are non-zero only for processes involving $J/\psi$ or $\Upsilon$ production.
}
\label{fig:diags}
\end{figure}

\section{Perturbative QCD amplitude for $e^+e^-\to q\bar q+\gamma$}\label{sec:amplitude}
We consider four tree-level diagrams given in Fig.~\ref{fig:diags}. In the first two diagrams the heavy quark-antiquark pair $q\bar q$ is produced through the annihilation of virtual $Z^0$-boson, while the last two diagrams involve fragmentation of virtual photon~$\gamma^*$ directly into $q\bar q$. There are also exist additional tree-level diagrams: for example, the ones obtained by interchanging $Z^0$-boson and $\gamma^*$ in Fig.~\ref{fig:diags}, but their contribution is found to be negligible in the vicinity of $\sqrt s=M_Z$~\cite{chen2013}.
The $\gamma^*$-fragmentation diagrams are important for $J/\psi$ and $\Upsilon$ production, but their contribution is zero for other $S$- and $P$-wave quarkonium states. Also, we note that the diagrams in Fig.~\ref{fig:diags} have trivial color factor~$\delta^{ab}$, so there are no contributions from octet color states to $e^+e^-\to H+\gamma$ at the considered order~$\alpha^3$.

The amplitude reads as
\bga
\label{eq:mp-gen}
\mathcal M_0=\frac{e^3\sqrt3|\mathcal Q|\varepsilon_{\gamma\nu}^*(Q)}{4\sin\theta_w\cos\theta_w} \,\bar v(k_2)\gamma_\mu({1-\gamma_5}-4\sin^2\theta_w)u(k_1)
\\ \times 
\Tr \Pi_{S}\Bigl[\gamma^\nu\frac{m+\hat p_1+\hat Q}{(p_1+Q)^2-m^2}L_\beta +L_\beta\frac{m-\hat p_2-\hat Q}{(p_2+Q)^2-m^2}\gamma^\nu\Bigr]D_Z^{\mu\beta}(P+Q)
\\
-e^3\sqrt3\mathcal Q\,\varepsilon_{\gamma\nu}^*(Q) \, \bar v(k_2)\Bigl[\gamma_\mu\frac{m_e-\hat k_2+\hat P}{(k_2-P)^2-m_e^2}\gamma^\nu
+\gamma^\nu\frac{m_e-\hat k_2+\hat Q}{(k_2-Q)^2-m_e^2}\gamma_\mu \Bigr]
 u(k_1)\Tr[\Pi_S\gamma_\beta]D_{\gamma}^{\mu\beta}(P),
\\
L_\beta=\frac{\gamma_\beta({1-\gamma_5}-4|\mathcal Q|\sin^2\theta_w)}{4\sin\theta_w\cos\theta_w},
\ega
where $e=\sqrt{4\pi\alpha}$, $\mathcal Q$ is heavy quark charge, $m$ is heavy quark mass, $m_e$ is electron mass, $\theta_w$ is weak mixing angle, $D_{\gamma(Z)}$ is virtual photon~$\gamma^*$ ($Z^0$-boson) propagator, and $\varepsilon_\gamma(Q)$ is polarization vector of the outgoing photon~$\gamma$.
The hat symbol denotes contraction with Dirac gamma-matrices.
As shown in Fig.~\ref{fig:diags}, $k_{1,2}$ is four-momentum of initial $e^\mp$ particle, $P$ and $Q$ are four-momenta of resulting quarkonium~$H$ and photon~$\gamma$, respectively. The total four-momentum of the produced $q\bar q$ pair is set to $p_1+p_2=P$, so that the momenta $p_{1,2}$ of quark and antiquark can be decomposed as $p_{1,2}=\frac12P\pm p$, with $P=(2\epsilon_{\boldsymbol p},\boldsymbol0)$ and $p=(0,\boldsymbol p)$ in the meson rest frame. The relativistic projection operators on the given spin state have the following form~\cite{bodwin2002}:
\bga
\Pi_{S=0,1}=\frac{(-1)^S}{4\sqrt2\epsilon_{\boldsymbol p}(\epsilon_{\boldsymbol p}+m)}(\hat p_2-m)(\gamma_5\delta_{S,0}+\hat\varepsilon^*(P)\delta_{S,1})[{\hat P}+{2\epsilon_{\boldsymbol p}}](\hat p_1+m)
\ega
where $\varepsilon(P)$ is polarization vector for spin-triplet states, $\varepsilon(P)\cdot P=0$.

\section{Angular integration}
\label{sec:ang-int}
Before being substituted in Eq.~\eqref{eq:nrqcd-amp}, the amplitude~\eqref{eq:mp-gen} has to be projected onto the required orbital-angular-momentum
state of the specific hadron~$H$. This is achieved by angular integration over directions of the relative momentum~$\boldsymbol p$ of heavy quark-antiquark pair $q\bar q$:
\bga
\mathcal M_{L=0}=\frac{1}{4\pi}\int \mathcal M_0 \,d\Omega, \qquad
\mathcal M_{L=1}=\frac{i\sqrt3}{4\pi}\int \frac{\boldsymbol \epsilon^* \boldsymbol p}{|\boldsymbol p |} \mathcal M_0 \,d\Omega,
\ega
where $\boldsymbol \epsilon$ is orbital polarization vector for $P$-wave quarkonium.

The integration of the amplitude~\eqref{eq:mp-gen} can be performed analytically. We choose reference frame with the $z$-axis along the momentum direction~$\boldsymbol P$ of the produced quarkonium:
\bga
\label{eq:kinematics}
k_{1,2}=\frac12({\sqrt s},\pm\sqrt{s-4m_e^2}\,\sin\vartheta,0,\pm\sqrt{s-4m_e^2}\,\cos\vartheta),\\
P=\frac1{2\sqrt s}(s+4\epsilon^2_{\boldsymbol p},0,0,s-4\epsilon^2_{\boldsymbol p}),\\
Q=\frac1{2\sqrt s}(s-4\epsilon^2_{\boldsymbol p},0,0,-s+4\epsilon^2_{\boldsymbol p}).
\ega
The general form of the relative three-momentum is $\boldsymbol p=|\boldsymbol p |(\sin\theta\cos\phi,\sin\theta\sin\phi,\cos\theta)$ and $d\Omega=\sin\theta\,d\theta d\phi$.
The amplitude~\eqref{eq:mp-gen} contains up to the third power of~$p_\mu$, which is
\bga
\label{eq:p-def}
p=\frac{|\boldsymbol p |}{4\epsilon_{\boldsymbol p}{\sqrt s}}((s-4\epsilon^2_{\boldsymbol p})\cos\theta,\sin\theta\cos\phi,\sin\theta\sin\phi,(s+4\epsilon^2_{\boldsymbol p})\cos\theta)
\ega
in the considered reference frame. Then, we immediately obtain
\bga
\frac{1}{4\pi}\int\frac{d\Omega}{(p_{1,2}+Q)^2-m^2}=\frac{2\epsilon_{\boldsymbol p}\,\mathcal C}{|\boldsymbol p|(s-4\epsilon_{\boldsymbol p}^2)},
\\
\frac{1}{4\pi}\int\frac{p_\mu\,d\Omega}{(p_{1,2}+Q)^2-m^2}=\frac{\pm1}{s-4\epsilon_{\boldsymbol p}^2}(1-\xi\mathcal C)\tilde P_\mu,
\\
\frac{1}{4\pi}\int\frac{p_\mu p_\nu\,d\Omega}{(p_{1,2}+Q)^2-m^2}=\frac{|\boldsymbol p|\epsilon_{\boldsymbol p}}{s-4\epsilon_{\boldsymbol p}^2}(2\xi(1-\xi\mathcal C)P_{\mu\nu}-(3\xi+\mathcal C-3\xi^2\mathcal C)G_{\mu\nu}),
\\
\frac{1}{4\pi}\int\frac{p_\mu p_\nu p_\lambda \,d\Omega}{(p_{1,2}+Q)^2-m^2}=\frac{\mp{\boldsymbol p}^2}{18(s-4\epsilon_{\boldsymbol p}^2)}[2(1+3\xi^2-3\xi^3\mathcal C)(\tilde P_\mu P_{\nu\lambda}+\tilde P_\nu P_{\mu\lambda}+\tilde P_\lambda P_{\mu\nu})
\\
+(4-15\xi^2-9\xi\mathcal C+15\xi^3\mathcal C)(\tilde P_\mu G_{\nu\lambda}+\tilde P_\nu G_{\mu\lambda}+\tilde P_\lambda G_{\mu\nu})],
\ega
where we additionaly introduced
\bgas
\tilde P=\frac1{2\sqrt s}(s-4\epsilon^2_{\boldsymbol p},0,0,s+4\epsilon^2_{\boldsymbol p}),\quad
P_{\mu\nu}=g_{\mu\nu}-\frac{P_\mu P_\nu}{4\epsilon^2_{\boldsymbol p}},\quad
G_{\mu\nu}=\diag(0,-1,-1,0),\\
\xi(\boldsymbol p)=\epsilon_{\boldsymbol p}/|{\boldsymbol p}|, \qquad \mathcal C(\boldsymbol p)=\coth^{-1}\xi.
\egas
The analogous but more lengthy expressions can also be written for the $P$-wave case. The projection of the $P$-wave states onto the required total orbital momentum~$J$ is completed by the relations
\bga
\sum_{S_z,L_z}\langle1S_z;1L_z|JJ_z\rangle\,\varepsilon_{\mu}^*(S_z)\,\epsilon_{\nu}^*(L_z)=
\begin{cases}
\dfrac1{\sqrt3}P_{\mu\nu}, &J=0,\\
\dfrac{i\,{e_{\mu\nu}}^{\lambda\omega}}{2\sqrt2\,\epsilon_{\boldsymbol p}}P_\lambda\varepsilon_\omega^*(J_z), &J=1,\\
\varepsilon_{\mu\nu}^*(J_z), &J=2.
\end{cases}
\ega
Finally, for summation over polarizations of the outgoing $S$- or $P$-wave quarkonium, we use
\bga
\sum_{J_z}\varepsilon_\mu\varepsilon_\nu^*=P_{\mu\nu},\\
\sum_{J_z}\varepsilon_{\mu\nu}\varepsilon_{\lambda\omega}^*=\frac12(P_{\mu\lambda}P_{\nu\omega}+P_{\mu\omega}P_{\nu\lambda})-\frac13P_{\mu\nu}P_{\lambda\omega}.
\ega

\section{Resummed NRQCD cross sections and VMD result for $e^+e^-\to J/\psi(\Upsilon)+\gamma$}\label{sec:jpsi}
We present the cross section of spin-triplet $^3S_1$ quarkonium production as the sum of two terms
\be
\label{eq:cs-3s1}
\sigma[{^3S_1}]=\sigma_Z+\sigma_{\gamma^*},
\ee
where $\sigma_Z$ originates from virtual $Z^0$-boson diagrams in Fig.~\ref{fig:diags},
and $\sigma_{\gamma^*}$ contains only the virtual photon contributions.
The cross term between these two sets of diagrams is negligible near $\sqrt s=M_Z$, so we do not consider it here.
The resummed NRQCD result for the $Z^0$-boson channel is
\bga
\label{eq:sigma-z}
\frac{d\sigma_Z[{^3S_1}]}{d\cos\theta}=
3\pi\alpha^3(s-M^2) M |R(0)|^2
\frac{\mathcal Q^2(1-4s^2_w+8s^4_w)}
{128s^2\,s^4_wc^4_w((s-M_Z^2)^2+M_Z^2\Gamma_Z^2)}
\\
\times
\Biggl[
\frac{
2\bigl(|\boldsymbol p|\epsilon_{\boldsymbol p}
(3ms+2s\epsilon_{\boldsymbol p}-4m\epsilon^2_{\boldsymbol p})-m^2(ms+4m\epsilon^2_{\boldsymbol p}+8\epsilon^3_{\boldsymbol p})\coth^{-1}\bigl[\frac{\epsilon_{\boldsymbol p}}{|{\boldsymbol p}|}\bigr]\bigr)^2
}
{\epsilon^2_{\boldsymbol p}(m+\epsilon_{\boldsymbol p})^2(s-4\epsilon^2_{\boldsymbol p})^2\boldsymbol p^2}(1-\cos^2\theta)
\\+
\frac{
s\bigl(|\boldsymbol p|(s+4m\epsilon_{\boldsymbol p})+m(s-4m^2-4m\epsilon_{\boldsymbol p}-4\epsilon^2_{\boldsymbol p})\coth^{-1}\bigl[\frac{\epsilon_{\boldsymbol p}}{|{\boldsymbol p}|}\bigr]\bigr)^2
}
{(m+\epsilon_{\boldsymbol p})^2(s-4\epsilon^2_{\boldsymbol p})^2\boldsymbol p^2}(1+\cos^2\theta)
\Biggr]_{\boldsymbol p^{2}={\langle {\boldsymbol p}^{2} \rangle}},
\ega
where $s(c)_w=\sin(\cos)\theta_w$, $M_Z$ and $\Gamma_Z$ is the $Z^0$-boson mass and total decay width, respectively,
and $\epsilon_{\boldsymbol p}=\sqrt{m^2+\boldsymbol p^2}$ is the energy of heavy $b$- or $c$-quark with the mass~$m$.
The factors containing $J/\psi$ or $\Upsilon$ mass~$M$ come from relativistic normalization of the NRQCD amplitude~\eqref{eq:nrqcd-gen} and from the two-body phase space.
As a result of resummation, the cross section~\eqref{eq:sigma-z} is an even function of the heavy quark relative momentum~$|{\boldsymbol p}|$, which has to be taken at the value $\boldsymbol p^{2}={\langle {\boldsymbol p}^{2} \rangle}$.
The correction of relative order $O(v^2)$ to the completely nonrelativistic cross section calculated in NRQCD can be readily obtained from Eq.~\eqref{eq:sigma-z} by simple linear expansion in~${\langle {\boldsymbol p}^{2} \rangle}$ (but this is generally not true for relativistic corrections of higher orders).

Analogously, for the second term in~\eqref{eq:cs-3s1} we have
\bga
\label{eq:sigma-gstar-nrqcd}
\frac{d\sigma^\text{NRQCD}_{\gamma^*}[{^3S_1}]}{d\cos\theta}
=\frac{\mathcal Q^2}{6s^3}\pi\alpha^3(s-M^2) M |R(0)|^2
\biggl[
\frac{(m+2\epsilon_{\boldsymbol p})^2}
{\epsilon^6_{\boldsymbol p}(s-4\epsilon^2_{\boldsymbol p})^2(1-(1-{4m_e^2}/s)\cos^2\theta)^2}
\\
\times\bigl(
s[8s\epsilon^2_{\boldsymbol p}\sin^4\theta+(s^2+16\epsilon^4_{\boldsymbol p})(1-\cos^4\theta)]+8m_e^2[s(s-8\epsilon^2_{\boldsymbol p})\sin^2\theta+(s-4\epsilon^2_{\boldsymbol p})^2\cos^4\theta]
\bigr)\biggr]_{\boldsymbol p^{2}={\langle {\boldsymbol p}^{2} \rangle}}.
\ega
Note that the electron mass~$m_e$ is neglected in the cross sections except for numerically important cases.

The last two diagrams in Fig.~\ref{fig:diags} describe fragmentation of the virtual photon~$\gamma^*$ into vector quarkonium.
So, their contribution can also be calculated within the vector-meson-dominance formalism~\cite{bodwin-vmd,bodwin-resum,fan,sang15}.
We calculate the amplitude of $e^+e^-\to\gamma^*+\gamma$ and then couple $\gamma^*$ to vector quarkonium with the coupling constant
\bes
g_V=\sqrt{\frac{3M^3}{4\pi\alpha^2}\,\Gamma[{H\to e^+e^-}]}\,,
\ees
where $\Gamma[{H\to e^+e^-}]$ is the corresponding decay rate of $J/\psi$ or $\Upsilon$.
For this calculation the physical meson mass can be used in the kinematical conditions~\eqref{eq:kinematics}:
\bgas
P=\frac1{2\sqrt s}(s+M^2,0,0,s-M^2),\\
Q=\frac1{2\sqrt s}(s-M^2,0,0,-s+M^2).
\egas
Then, we finally obtain
\bga
\label{eq:sigma-gstar-vmd}
\frac{d\sigma^\text{VMD}_{\gamma^*}[{^3S_1}]}{d\cos\theta}
=\frac{6\pi\alpha\,\Gamma[{H\to e^+e^-}]}{s^3M(s-M^2)(1-(1-{4m_e^2}/s)\cos^2\theta)^2}
\\
\times
\bigl(
s[(s+M^2)^2-4sM^2\cos^2\theta-(s-M^2)^2\cos^4\theta]+8m_e^2[s(s-2M^2)\sin^2\theta+(s-M^2)^2\cos^4\theta]
\bigr).
\ega
The photon-fragmentation diagrams do not contribute to production of the other $S$- and $P$-wave quarkonium states.
The explicit form of cross sections for these processes along with the angularly integrated expressions~\eqref{eq:sigma-z}--\eqref{eq:sigma-gstar-vmd} can be found in Appendix.

\section{Numerical results}\label{sec:numerical}
For particle masses, decay widths and weak coupling angle we use~\cite{pdg}:
$m_e=5.1\times10^{-4}$,
$M_Z=91.188$,
$\Gamma_Z=2.495$,
$\Gamma[{J/\psi\to e^+e^-}]=(5.55\pm0.16)\times10^{-6}$,
$\Gamma[{\Upsilon\to e^+e^-}]=(1.34\pm0.02)\times10^{-6}$~GeV,
$\sin^2\theta_w=0.231$.
For fine structure constant we take
$\alpha(M_Z)=1/128.9$,
$\alpha(M_\Upsilon)=1/131.1$,
$\alpha(M_{J/\psi})=1/132.6$, and
$\alpha(0)=1/137$
in dependence on the actual momentum transfer in vertices in Fig.~\ref{fig:diags}.
Quarkonium masses are given in the Appendix, Eq.~\eqref{quarkonium_masses}.
For quark masses we take $m_c=1.5\pm0.1$ and $m_b=4.6\pm0.1$~GeV.
We adopt the following values of quarkonium NRQCD matrix elements from Refs.~\cite{bodwin2008,chung2008,chung2010,bodwin16,eichten96}:
\bga
\label{eq:me-nums}
|R(0)|^2_{(c\bar c)}=0.91\pm0.14~\text{GeV}^3, \quad
|R'(0)|^2_{(c\bar c)}=0.042^{+0.030}_{-0.020}~\text{GeV}^5, \quad
{\langle {\boldsymbol p}^{2} \rangle}_{(c\bar c)}=0.44\pm0.14~\text{GeV}^2,
\\
|R(0)|^2_{(b\bar b)}=6.43\pm0.44~\text{GeV}^3, \qquad
|R'(0)|^2_{(b\bar b)}=1.42~\text{GeV}^5, \qquad
{\langle {\boldsymbol p}^{2} \rangle}_{(b\bar b)}=-0.19\pm0.22~\text{GeV}^2.
\ega
We do not distinguish different spin states in~\eqref{eq:me-nums} since the corresponding corrections are almost order of magnitude smaller than  main uncertainties and we also use the same value of $\langle {\boldsymbol p}^{2} \rangle$ for $S$- and $P$-wave charmonium or bottomonium states.
Note that the central value ${\langle {\boldsymbol p}^{2} \rangle}_{(b\bar b)}=-0.19~\text{GeV}^2$ extracted in Ref.~\cite{chung2010} 
from the electronic width of $\Upsilon$ differs drastically from the estimate 
${\langle {\boldsymbol p}^{2} \rangle}_{(b\bar b)}=1.21~\text{GeV}^2$
based on the Gremm--Kapustin relation~\cite{g-k,chen12}.

\begin{figure}[t!]
\begin{center}
\includegraphics{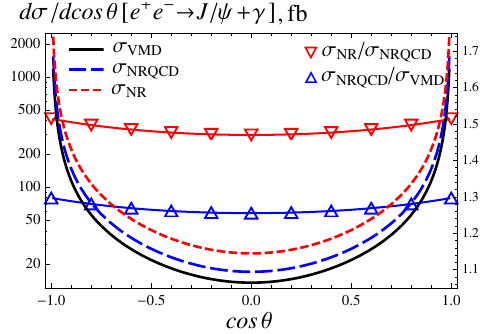}\qquad\includegraphics{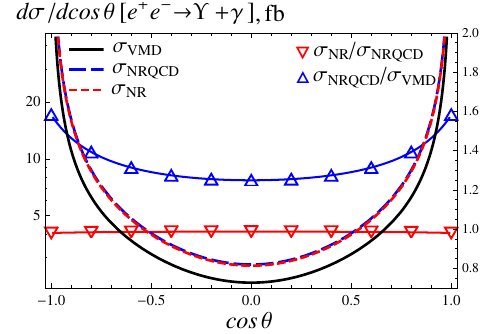}
\end{center}
\caption{Angular distributions of $J/\psi$ (left) and $\Upsilon$ (right) produced in the process $e^+e^-\to J/\psi(\Upsilon)+\gamma$ at the energy $\sqrt s =M_Z$. $\sigma_\text{VMD}$ and $\sigma_\text{NRQCD}$ denotes the total cross section~\eqref{eq:cs-3s1} with VMD~\eqref{eq:sigma-gstar-vmd} or NRQCD~\eqref{eq:sigma-gstar-nrqcd} treatment of the photon-fragmentation contributions, respectively. $\sigma_\text{NR}=\sigma_\text{NRQCD}[\langle {\boldsymbol p}^{2} \rangle=0]$ is the nonrelativistic limit of NRQCD cross section. The scale at the right edge of figures corresponds to the ratios of cross sections.
}
\label{fig:cos-theta} 
\end{figure}

\begin{figure}[t!]
\begin{center}
\includegraphics{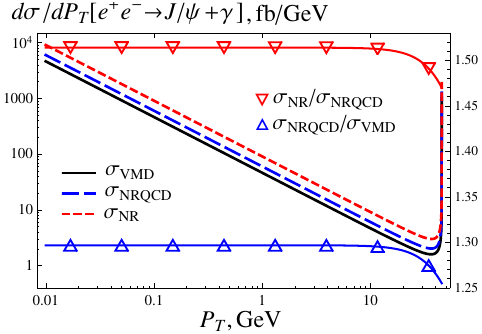}\qquad\includegraphics{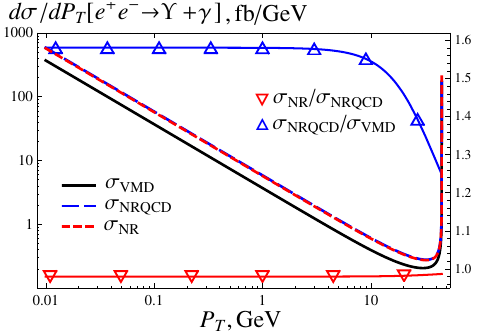}
\end{center}
\caption{Transverse momentum $P_T$ distributions of $J/\psi$ (left) and $\Upsilon$ (right) produced in the process $e^+e^-\to J/\psi(\Upsilon)+\gamma$ at the energy $\sqrt s =M_Z$. The notations as for Fig.~\ref{fig:cos-theta}.
}
\label{fig:pt-3s1}
\end{figure}

The numerical results for charmonium and bottomonium production are given in Tables~\ref{tbl1} and~\ref{tbl2}, respectively.
The second column contains relativistic cross sections, while in the third column relative contribution of relativistic corrections with respect to the nonrelativistic result is given.
For comparison, we also list in Tables~\ref{tbl1} and~\ref{tbl2} relative contributions of one-loop QCD corrections to nonrelativistic cross sections obtained in Refs~\cite{chen2014,sun2014}.
For $J/\psi$ and $\Upsilon$ the calculation was performed with both NRQCD~\eqref{eq:sigma-gstar-nrqcd} and VMD~\eqref{eq:sigma-gstar-vmd} results for $\sigma_{\gamma^*}$ in Eq.~\eqref{eq:cs-3s1}.
This contribution almost completely determines $J/\psi$ and $\Upsilon$ production, so in the VMD case the relativistic corrections to the total cross section are negligible, since the VMD result~\eqref{eq:sigma-gstar-vmd} already contains relativistic and also QCD corrections to the photon fragmentation.
If all contributions are treated within NRQCD, the total relativistic corrections to $\sigma[e^+e^-\to J/\psi+\gamma]$ are large and negative, shifting the nonrelativistic result towards the VMD prediction.
There is no complete one-loop QCD analysis for this case in the literature, but the analogous calculation for $\Upsilon$ production~\cite{sun2014} shows that QCD corrections are also negative, in agreement with the VMD result. The relativistic corrections to $\sigma[e^+e^-\to \Upsilon+\gamma]$ are negligible due to the smallness of ${\langle {\boldsymbol p}^{2} \rangle}_{(b\bar b)}$.
In Figs.~\ref{fig:cos-theta} and~\ref{fig:pt-3s1} we plot angular and transverse momentum $P_T$ distributions for $J/\psi$ and $\Upsilon$ cross sections. For $J/\psi$, NRQCD and VMD corrections do not significantly change the shape of both distributions in the considered regions, so that their action on the nonrelativistic cross section can be approximated by the lowering factors $K_\text{NRQCD}\approx1.5$ and $K_\text{VMD}\approx2.0$, respectively.
The VMD contribution to the nonrelativitistic $\Upsilon$ cross section is not so uniform: $\sigma^\Upsilon_\text{NR}/\sigma^\Upsilon_\text{VMD}\approx1.3$ in the interval $|\!\cos\theta|<0.5$ and growths up to $\sigma^\Upsilon_\text{NR}/\sigma^\Upsilon_\text{VMD}\approx1.6$ at the both ends $\cos\theta=\pm1$ of angular distribution range.
Analogously, the considered ratio stays almost constant for transverse momenta $P_T\lesssim5$~GeV, and then starts to decrease from 1.6 down to $\sim\!1.3$ at the highest $P_T$.

Finally, in Table~\ref{tbl3} we compare resummed results for charmonium cross sections with their truncated form, where only leading correction in $v^2$ is preserved. The later corresponds to the formal order of validity of the generalized Gremm--Kapustin relation~\eqref{eq:nrqcd-ggkr}, and so of the resummed cross sections, as discussed in Section~\ref{sec:nrqcd_resum}. Table~\ref{tbl3} shows that, in general, leading corrections agree well with the whole resummed result, difference not exceeding 5--7\%. The only exception is $J/\psi$, where linear corrections give the cross section an additional 12\% decrease.

\begin{table}[t!]
\caption{Cross sections of charmonium~$H$ production in the process $e^+e^-\to H+\gamma$ at the energy $\sqrt s =M_Z$. $\sigma$ is the relativistic resummed result, $\sigma_\text{NR}=\sigma[\langle {\boldsymbol p}^{2} \rangle=0]$ is the cross section in nonrelativistic approximation, $\sigma_\text{NLO}$ is the nonrelativistic cross section with one-loop QCD corrections included. $\Delta\sigma=\sigma-\sigma_\text{NR}$ and $\Delta\sigma_\text{NLO}=\sigma_\text{NLO}-\sigma_\text{NR}$.
\label{tbl1}}
\bigskip
\begin{center}
\begin{tabular}{lccc}
\hline
$H$ & $\sigma$, fb &  $\Delta\sigma/\sigma_\text{NR}$, \% & $\Delta\sigma_\text{NLO}/\sigma_\text{NR}$, \%~\cite{chen2014} \\[1pt]    \hline
\\ [-12pt]
$J/\psi$ NRQCD & $700^{+223}_{-188}$ & $-34$ & --- \\
$J/\psi$ VMD & $541\pm16$ & $-0.1$ & 0.3 \\
$\eta_c$ & $0.64\pm0.13$ & $-21$ & 19 \\
$\chi_{c0}$ & $0.014^{+0.011}_{-0.007}$ & $-28$ & $1.1$ \\[3pt]
$\chi_{c1}$ & $0.089^{+0.068}_{-0.047}$ & $-25$ & $15$ \\[3pt]
$\chi_{c2}$ & $0.032^{+0.025}_{-0.017}$ & $-20$ & $-76$ \\[3pt]
$h_c$ & $0.29^{+0.22}_{-0.15}$ & $-28$ & $-66$ \\ \hline
\end{tabular}
\end{center}
\end{table}

\begin{table}[t!]
\caption{Cross sections of bottomonium~$H$ production in the process $e^+e^-\to H+\gamma$ at the energy $\sqrt s =M_Z$.
The notations as for Table~\ref{tbl1}.
\label{tbl2}}
\bigskip
\begin{center}
\begin{tabular}{lccc}
\hline
$H$ & $\sigma$, fb &  $\Delta\sigma/\sigma_\text{NR}$, \% & $\Delta\sigma_\text{NLO}/\sigma_\text{NR}$, \%~\cite{sun2014} \\[1pt]    \hline
\\ [-12pt]
$\Upsilon$ NRQCD & $71\pm8$ & $2$ & $-30$ \\
$\Upsilon$ VMD & $46\pm1$ & $0.04$ & 0.2 \\
$\eta_b$ & $1.58\pm0.13$ & 1.2 & $-1.6$ \\
$\chi_{b0}$ & $0.017\pm0.002$ & 1.7 & $4.5$ \\
$\chi_{b1}$ & $0.11\pm0.01$ & $1.4$ & $-5$ \\
$\chi_{b2}$ & $0.036\pm0.003$ & 1.1 & $-52$ \\
$h_b$ & $0.11\pm0.01$ & $1.6$ & $-44$ \\ \hline
\end{tabular}
\end{center}
\end{table}

\begin{figure}[t!]
\begin{center}
\includegraphics{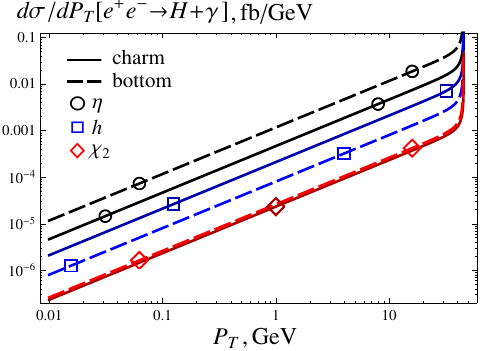}\qquad\includegraphics{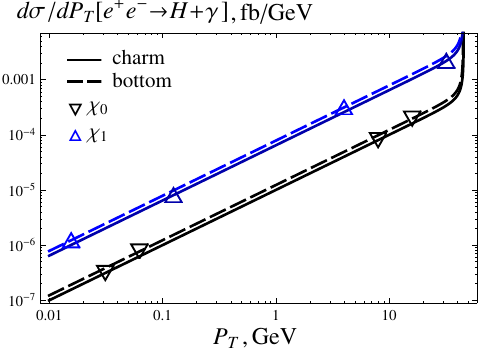}
\end{center}
\caption{Transverse momentum $P_T$ distributions of heavy quarkonium~$H$ produced in the process $e^+e^-\to H+\gamma$ at the energy $\sqrt s =M_Z$. The dashed and solid lines correspond to bottomonium and charmonium states, respectively.
}
\label{fig:pt-others} 
\end{figure}

\begin{table}[t!]
\caption{Charmonium cross sections $\sigma[e^+e^-\to H+\gamma]$ at $\sqrt s =M_Z$ in nonrelativistic approximation and with relativistic corrections.
$\sigma_\text{NR}=\sigma[\langle {\boldsymbol p}^{2} \rangle=0]$ is the nonrelativistic limit, $\sigma_\text{RESUM}$ is the resummed cross section, which can be expanded in heavy quark velocity~$v$ as $\sigma_\text{RESUM}=\sigma_\text{NR}+[d\sigma/d\langle {\boldsymbol p}^{2} \rangle]_{\langle {\boldsymbol p}^{2} \rangle=0}\langle {\boldsymbol p}^{2} \rangle+\mathcal O_\sigma(v^4)\equiv \sigma_\text{LIN}+\mathcal O_\sigma(v^4)$, where $\sigma_\text{LIN}$ is the linear (in $\langle {\boldsymbol p}^{2} \rangle$) approximation to the relativistic cross section~$\sigma$ and $\mathcal O_\sigma(v^4)$ is the residue of $v^2$-expansion estimated on the basis of Eq.~\eqref{eq:nrqcd-ggkr}.
$\Delta\sigma=\sigma_\text{RESUM}-\sigma_\text{NR}$ and $\Delta\sigma_\text{LIN}=\sigma_\text{LIN}-\sigma_\text{NR}$.
For $J/\psi$, the NRQCD result~\eqref{eq:sigma-gstar-nrqcd} is used.
\label{tbl3}}
\bigskip
\begin{center}
\begin{tabular}{lcccccc}
\hline
$H$ & $\sigma_\text{NR}$, fb &  $\sigma_\text{LIN}$, fb & $\sigma_\text{RESUM}$, fb & $\Delta\sigma/\sigma_\text{NR}$, \% & $\Delta\sigma_\text{LIN}/\sigma_\text{NR}$, \% & $\mathcal O_\sigma(v^4)/\sigma_\text{NR}$, \% \\[1pt]    \hline
\\ [-12pt]
$J/\psi$ & 1058 & 578 & 700 & $-34$ & $-45$ & $12$ \\[1pt]
$\eta_c$ & $0.82$ & $0.61$ & $0.64$ & $-21$ & $-26$ & 5 \\[1pt]
$\chi_{c0}$ & $0.019$ & $0.012$ & $0.014$ & $-28$ & $-35$ & 7 \\[1pt]
$\chi_{c1}$ & $0.12$ & $0.082$ & $0.089$ & $-25$ & $-31$ & 6 \\[1pt]
$\chi_{c2}$ & $0.040$ & $0.031$ & $0.032$ & $-20$ & $-23$ & 4 \\[1pt]
$h_c$ & $0.40$ & $0.26$ & $0.29$ & $-28$ & $-35$ & 7 \\ \hline
\end{tabular}
\end{center}
\end{table}

\section{Summary}
In this paper, the NRQCD resummation of relativistic corrections to heavy $\text{quarkonium}+\gamma$ production at $Z$ factory is performed at LO in $\alpha_s$, and analytical expressions for cross sections are given. The $Z^{0}$-boson channel is dominant for the production, with exception of $J/\psi(\Upsilon)$. There, as it shown by VMD analysis, photon fragmentation channel gives the main contribution, with both relativistic and NLO $\alpha_s$ corrections to the $Z^{0}$ channel becoming insignificant. Still, when treated purely within NRQCD, this process also follows the general trend for another charmonium states, with relativistic corrections having the same (or even much larger, as in the case of $\chi_{c0}$) significance as those in $\alpha_s$.

The relativistic corrections to all considered charmonium processes are negative, decreasing the cross sections 20--30\%, with the largest contribution of $-34\%$ in the case of NRQCD calculation for $J/\psi$. The significant drop of $J/\psi+\gamma$ cross section is also confirmed by VMD result, effectively including corrections of both types, which is almost 50\% lower than non-relativistic LO $\alpha_s$ NRQCD prediction.
As for bottomonium production, the VMD result for $\Upsilon$ ($-35$\%) is again in agreement with $-30$\% of NLO $\alpha_s$ corrections reported in~\cite{sun2014}. The relativistic corrections to bottomonium processes are positive but generally negligible, not exceeding 2\% (though, they are comparable with NLO QCD corrections for $\eta_b$, $\chi_{b0}$, and $\chi_{b1}$, which are also small), as it is expected due to much heavier quark mass and smaller value of quark relative velocity.

\renewcommand{\theequation}{A.\arabic{equation}}
\setcounter{equation}{0}
\section*{Appendix}
Quarkonium masses used for calculation:
\bga
\label{quarkonium_masses}
M_{J/\psi}=3.097, \quad
M_{\eta c}=2.983,\quad
M_{\chi{c0}}=3.415, \quad
M_{\chi{c1}}=3.511, \quad
M_{\chi{c2}}=3.556,\quad
M_{hc}=3.525,\\
M_\Upsilon=9.460, \quad
M_{\eta b}=9.399, \quad
M_{\chi{b0}}=9.859,\quad
M_{\chi{b1}}=9.893,\quad
M_{\chi{b2}}=9.912,\quad
M_{hb}=9.899~\text{GeV}.
\ega

Explicit form of differential and integrated cross sections:
\bgas
\sigma[{^1S_0}]=
\pi\alpha^3m^2(s-M^2) M |R(0)|^2
\frac{\mathcal Q^2(1-4|\mathcal Q|s^2_w)^2(1-4s^2_w+8s^4_w)}
{16s\,s^4_wc^4_w((s-M_Z^2)^2+M_Z^2\Gamma_Z^2)}
\biggl[
\frac{\coth^{-1}\bigl[\frac{\epsilon_{\boldsymbol p}}{|{\boldsymbol p}|}\bigr]}{\epsilon_{\boldsymbol p}|\boldsymbol p|}
\biggr]^2_{\boldsymbol p^{2}={\langle {\boldsymbol p}^{2} \rangle}},\\
\frac{d\sigma[{^1S_0}]}{d\cos\theta}=\frac38(1+\cos^2\theta)\,\sigma[{^1S_0}].
\egas

\vspace{1em}

\bgas
\sigma[{^3P_0}]=
9\pi\alpha^3m^2(s-M^2) M |R(0)|^2
\frac{\mathcal Q^2(1-4|\mathcal Q|s^2_w)^2(1-4s^2_w+8s^4_w)}
{16s\,s^4_wc^4_w((s-M_Z^2)^2+M_Z^2\Gamma_Z^2)}\\
\times
\Biggl[
\frac{s|\boldsymbol p|-\epsilon_{\boldsymbol p}(s-4{\boldsymbol p}^2)\coth^{-1}\bigl[\frac{\epsilon_{\boldsymbol p}}{|{\boldsymbol p}|}\bigr]
}
{\epsilon_{\boldsymbol p}(s-4\epsilon^2_{\boldsymbol p})\boldsymbol p^3}
\Biggr]^2_{\boldsymbol p^{2}={\langle {\boldsymbol p}^{2} \rangle}},\\
\frac{d\sigma[{^3P_0}]}{d\cos\theta}=\frac38(1+\cos^2\theta)\,\sigma[{^3P_0}].
\egas

\vspace{1em}

\bgas
\sigma[{^3P_1}]=
27\pi\alpha^3(s-M^2) M |R(0)|^2
\frac{\mathcal Q^2(1-4|\mathcal Q|s^2_w)^2(1-4s^2_w+8s^4_w)}
{32\,s^4_wc^4_w((s-M_Z^2)^2+M_Z^2\Gamma_Z^2)}\\
\times
\Biggl[
\frac{({s+4\epsilon^2_{\boldsymbol p}})(\epsilon_{\boldsymbol p}|\boldsymbol p|-m^2\coth^{-1}\bigl[\frac{\epsilon_{\boldsymbol p}}{|{\boldsymbol p}|}\bigr])^2
}
{\epsilon^2_{\boldsymbol p}(s-4\epsilon^2_{\boldsymbol p})^2\boldsymbol p^6}
\Biggr]_{\boldsymbol p^{2}={\langle {\boldsymbol p}^{2} \rangle}},\\
\frac{d\sigma[{^3P_1}]}{d\cos\theta}=\frac38
\biggl[
\frac{s(1+\cos^2\theta)+8\epsilon^2_{\boldsymbol p}(1-\cos^2\theta)}{s+4\epsilon^2_{\boldsymbol p}}
\biggr]_{\boldsymbol p^{2}={\langle {\boldsymbol p}^{2} \rangle}}\sigma[{^3P_1}].
\egas

\vspace{1em}

\bgas
\sigma[{^1P_0}]=
27\pi\alpha^3 m^2(s-M^2)M |R(0)|^2
\frac{\mathcal Q^2(1-4s^2_w+8s^4_w)}
{16s^2\,s^4_wc^4_w((s-M_Z^2)^2+M_Z^2\Gamma_Z^2)}\\
\times
\Biggl[
\frac{(\epsilon_{\boldsymbol p}|\boldsymbol p|-m^2\coth^{-1}\bigl[\frac{\epsilon_{\boldsymbol p}}{|{\boldsymbol p}|}\bigr])^2}
{\epsilon^2_{\boldsymbol p}\boldsymbol p^6}
+\frac{s(|\boldsymbol p|-\epsilon_{\boldsymbol p}\coth^{-1}\bigl[\frac{\epsilon_{\boldsymbol p}}{|{\boldsymbol p}|}\bigr])^2
}
{\epsilon^2_{\boldsymbol p}\boldsymbol p^6}
\Biggr]_{\boldsymbol p^{2}={\langle {\boldsymbol p}^{2} \rangle}},
\egas

\bgas
\frac{d\sigma[{^1P_0}]}{d\cos\theta}=
81\pi\alpha^3 m^2(s-M^2)M |R(0)|^2
\frac{\mathcal Q^2(1-4s^2_w+8s^4_w)}
{128s^2\,s^4_wc^4_w((s-M_Z^2)^2+M_Z^2\Gamma_Z^2)}\\
\times
\Biggl[
2(1-\cos^2\theta)
\frac{(\epsilon_{\boldsymbol p}|\boldsymbol p|-m^2\coth^{-1}\bigl[\frac{\epsilon_{\boldsymbol p}}{|{\boldsymbol p}|}\bigr])^2}
{\epsilon^2_{\boldsymbol p}\boldsymbol p^6}
+(1+\cos^2\theta)\frac{s(|\boldsymbol p|-\epsilon_{\boldsymbol p}\coth^{-1}\bigl[\frac{\epsilon_{\boldsymbol p}}{|{\boldsymbol p}|}\bigr])^2
}
{\epsilon^2_{\boldsymbol p}\boldsymbol p^6}
\Biggr]_{\boldsymbol p^{2}={\langle {\boldsymbol p}^{2} \rangle}}.
\egas

\vspace{1em}

\bgas
\sigma^\text{NRQCD}_{\gamma^*}[{^3S_1}]=\frac{\mathcal Q^2}{3s^2}\pi\alpha^3(s-M^2) M |R(0)|^2(2\arctanh\sqrt{1-{4m_e^2}/s}-1)
\biggl[
\frac{(m+2\epsilon_{\boldsymbol p})^2(s^2+16\epsilon^4_{\boldsymbol p})}
{\epsilon^6_{\boldsymbol p}(s-4\epsilon^2_{\boldsymbol p})^2}
\biggr]_{\boldsymbol p^{2}={\langle {\boldsymbol p}^{2} \rangle}},
\egas

\bgas
\sigma^\text{VMD}_{\gamma^*}[{^3S_1}]
=\frac{12\pi\alpha\Gamma[{H\to l^+l^-}]}{s^2M(s-M^2)}(s^2+M^4)(2\arctanh\sqrt{1-{4m_e^2}/s}-1),
\egas

\bgas
\sigma_Z[{^3S_1}]=
\pi\alpha^3(s-M^2) M |R(0)|^2
\frac{\mathcal Q^2(1-4s^2_w+8s^4_w)}
{16s^2\,s^4_wc^4_w((s-M_Z^2)^2+M_Z^2\Gamma_Z^2)}
\\
\times
\Biggl[
\frac{
\bigl(|\boldsymbol p|\epsilon_{\boldsymbol p}
(3ms+2s\epsilon_{\boldsymbol p}-4m\epsilon^2_{\boldsymbol p})-m^2(ms+4m\epsilon^2_{\boldsymbol p}+8\epsilon^3_{\boldsymbol p})\coth^{-1}\bigl[\frac{\epsilon_{\boldsymbol p}}{|{\boldsymbol p}|}\bigr]\bigr)^2
}
{\epsilon^2_{\boldsymbol p}(m+\epsilon_{\boldsymbol p})^2(s-4\epsilon^2_{\boldsymbol p})^2\boldsymbol p^2}
\\+
\frac{
s\bigl(|\boldsymbol p|(s+4m\epsilon_{\boldsymbol p})+m(s-4m^2-4m\epsilon_{\boldsymbol p}-4\epsilon^2_{\boldsymbol p})\coth^{-1}\bigl[\frac{\epsilon_{\boldsymbol p}}{|{\boldsymbol p}|}\bigr]\bigr)^2
}
{(m+\epsilon_{\boldsymbol p})^2(s-4\epsilon^2_{\boldsymbol p})^2\boldsymbol p^2}
\Biggr]_{\boldsymbol p^{2}={\langle {\boldsymbol p}^{2} \rangle}}.
\egas

\vspace{1em}

\bgas
\frac{d\sigma[{^3P_2}]}{d\cos\theta}=
N_2|R'(0)|^2
\Biggl[
\frac{A_2(\boldsymbol p)+B_2(\boldsymbol p)(1+\cos^2\theta)
}
{(m+\epsilon_{\boldsymbol p})^2(s-4\epsilon^2_{\boldsymbol p})^2\boldsymbol p^6}
\Biggr]_{\boldsymbol p^{2}={\langle {\boldsymbol p}^{2} \rangle}},
\\
\sigma[{^3P_2}]=
\frac{2N_2}3|R'(0)|^2
\Biggl[
\frac{3A_2(\boldsymbol p)+4B_2(\boldsymbol p)
}
{(m+\epsilon_{\boldsymbol p})^2(s-4\epsilon^2_{\boldsymbol p})^2\boldsymbol p^6}
\Biggr]_{\boldsymbol p^{2}={\langle {\boldsymbol p}^{2} \rangle}},
\\
A_2(\boldsymbol p)=16s(m-\epsilon_{\boldsymbol p})^2(|\boldsymbol p|(4m+\epsilon_{\boldsymbol p})+3m^2\coth^{-1}[{\epsilon_{\boldsymbol p}}/{|{\boldsymbol p}|}])^2,\\
B_2(\boldsymbol p)=3\epsilon^2_{\boldsymbol p}\bigl(s|\boldsymbol p|+2m(s-4\epsilon^2_{\boldsymbol p})\coth^{-1}[{\epsilon_{\boldsymbol p}}/{|{\boldsymbol p}|}]\bigr)^2\\
-4\boldsymbol p^2\bigl(2m^4(16s-79\epsilon^2_{\boldsymbol p})-12m^3\epsilon_{\boldsymbol p}(s+28\epsilon_{\boldsymbol p}^2)
-m^2(3s^2-20s\epsilon_{\boldsymbol p}^2+260\epsilon_{\boldsymbol p}^4)+3m\epsilon_{\boldsymbol p}(s^2-2s\epsilon_{\boldsymbol p}^2-32\epsilon_{\boldsymbol p}^4)\\
+2\epsilon_{\boldsymbol p}^4(s-16\epsilon_{\boldsymbol p}^2)\bigr)
-6m^2|\boldsymbol p|\bigl(
88m^4\epsilon_{\boldsymbol p}+8m^3(3s+28\epsilon_{\boldsymbol p}^2)
-16m^2\epsilon_{\boldsymbol p}(3s-22\epsilon_{\boldsymbol p}^2)-2m(s^2+4s\epsilon_{\boldsymbol p}^2-176\epsilon_{\boldsymbol p}^4)\\
+5s^2\epsilon_{\boldsymbol p}-24s\epsilon_{\boldsymbol p}^3+160\epsilon_{\boldsymbol p}^5\bigr)\coth^{-1}[{\epsilon_{\boldsymbol p}}/{|{\boldsymbol p}|}]
+3m^3\bigl(40m^5+128m^4\epsilon_{\boldsymbol p}-16m^3(s-17\epsilon_{\boldsymbol p}^2)\\
+40m^2\epsilon_{\boldsymbol p}(s+8\epsilon_{\boldsymbol p}^2)+m(s^2-24s\epsilon_{\boldsymbol p}^2+224\epsilon_{\boldsymbol p}^4)-4\epsilon_{\boldsymbol p}(s-4\epsilon_{\boldsymbol p}^2)(s+8\epsilon_{\boldsymbol p}^2)\bigr)\coth^{-2}[{\epsilon_{\boldsymbol p}}/{|{\boldsymbol p}|}],\\
N_2=9\pi\alpha^3(s-M^2) M
\frac{\mathcal Q^2(1-4|\mathcal Q|s^2_w)^2(1-4s^2_w+8s^4_w)}
{256s\,s^4_wc^4_w((s-M_Z^2)^2+M_Z^2\Gamma_Z^2)}.
\egas

\end{document}